\documentclass[journal]{IEEEtran}
\usepackage{color}

\usepackage{cite}

\ifCLASSINFOpdf
  \usepackage[pdftex]{graphicx}
  \DeclareGraphicsExtensions{.pdf,.jpeg,.jpg}
\else
\fi
\usepackage{url}

\begin{document}
%
\title{Internet of Drones}
%
%
%

\author{Mirmojtaba~Gharibi,
        Raouf~Boutaba,~\IEEEmembership{Fellow,~IEEE,}
        and~Steven~L.~Waslander,~\IEEEmembership{Member,~IEEE}
\thanks{M. Gharibi and R. Boutaba are with D. Cheriton School
of Computer Science, University of Waterloo, Waterloo,
ON, N2L 3G1, Canada. Contacts: mgharibi@uwaterloo.ca and rboutaba@uwaterloo.ca respectively.}
\thanks{S. L. Waslander is with the Mechanical and Mechatronics Department, University of Waterloo, ON, N2L 3G1, Canada, stevenw@uwaterloo.ca}
}

\maketitle

\begin{abstract}
The Internet of Drones (IoD) is a layered network control architecture designed mainly for coordinating the access of unmanned aerial vehicles to controlled airspace, and providing navigation services between locations referred to as nodes. The IoD provides generic services for various drone applications such as package delivery, traffic surveillance, search and rescue and more. In this paper, we present a conceptual model of how such an architecture can be organized and we specify the features that an IoD system based on our architecture should implement. For doing so, we extract key concepts from three existing large scale networks, namely the air traffic control network, the cellular network, and the Internet and explore their connections to our novel architecture for drone traffic management. A simulation platform for IoD is being implemented which can be accessed from www.IoDnet.org in the future.
\end{abstract}


%
\IEEEpeerreviewmaketitle

\section{Introduction}
 

%
%
%
%
\IEEEPARstart{T}{he Internet of Drones}\footnote{Project's homepage: www.IoDnet.org} is an architecture designed for providing coordinated access to controlled airspace for unmanned aerial vehicles (UAVs), often referred to as drones. With the on-going miniaturization of sensors and processors and ubiquitous wireless connectivity, drones are finding many new uses in enhancing our way of life. There are many applications for drone technology, ranging from the on-demand package delivery, to traffic and wild life surveillance, inspection of infrastructure, search and rescue, agriculture, and cinematography. All these applications share a common need for both navigation and airspace management. In this work, we lay the architecture for generic services that can provide such a foundation for all current and future applications.

Among these applications, aerial package delivery will most urgently require a robust airspace allocation architecture, as it could result in many thousands of daily flights in the same geographic area, with many potential conflicts between drones navigating along similar or intersecting routes. The benefit to the global logistics network is clear, as drones could usher in a new era of on-demand delivery, and has been shown to be cost-competitive relative to ground-based delivery as well\cite{Raf14}, although longer haul transport clearly benefits from bundling onto larger transport vehicles.  Amazon states that about 83\% of their packages weigh below 2.5 kg \cite{CNN13}, a reasonable maximum payload for today's drones. Similarly, the average weight of packages delivered by Fedex is less than 5kg \cite{Fed15}.  In our opinion, this model can provide on-demand, inexpensive, and convenient access to the goods and items already in or near an urban area, including consumer goods, fast-food, medicine, and even on-demand groceries.

Despite a wave of drone package delivery prototype announcements (e.g. Matternet \cite{Rap13}, Amazon's prime air \cite{Ama13}, Google's project wing \cite{BBC14}, and DHL's Parcelcopter \cite{DHL14}), to the best of our knowledge, there has been no rigorous publication concerning the architecture of a drone-specific air traffic management system as the technology is still in its infancy. One good starting point is NASA's Unmanned Aerial System Traffic Management (UTM) project \cite{UTM15,UTM_fact15,Kop15}, which organized a symposium to begin preparations of a solution for low altitude traffic management to be proposed to the FAA. Related to this effort, both Amazon \cite{Ama_1_15,Ama_2_15} and Google \cite{Goo15} have published white papers which explore some of the strategies for managing the airspace and coordinating aerial vehicles through onboard system requirements such as ADS-B and V2V communication.  However, our contribution is to approach the drone airspace management problem by providing a universal architecture and a vocabulary of concepts to describe the IoD. In the future, different IoD systems can be developed based on it with their set of protocols and implementations of the features required by our IoD architecture. We suggest a possible operational model based on our architecture and we discuss the desired goals of the architecture and also the benefits that it provides as well as the subtleties that have to be addressed for any IoD system. 

Shortly after we published our preprint\cite{GBW16}, authors in \cite{DL16} published a preprint exploring some of the ideas pertaining to a UAV traffic network, called uNet. In uNet, instead of using a free-flight mode, similar to our architecture as will be explained in the section architecture, the airspace is divided into predefined routes. The authors argue that this provides for less reliance on advanced sense and avoid technologies and the ease of assigning conflict-free routes to the drones using the existing techniques. They consider use of sector-level uNets (sNets) where the traffic in each sector is under the authority of that particular uNet. We have a similar construct in our architecture with different zones where each zone is under authority of one or multiple Zone Service Provider (ZSP). However, one difference is that in our architecture, more than one ZSP can participate in managing the same zone. Furthermore, we take a systematic approach in defining the layers of the architecture as well as the features that have to be implemented for each layers.

\section{Relevant Networks}
For designing the architecture of the IoD, we study three distinct large scale network structures; namely air traffic control (ATC), cellular network, and the Internet. Each of these networks achieves some of the goals or functionalities we desire for the IoD. In each case, however, their conceptual architecture falls short of providing a thorough solution to the unique challenges of IoD. Hence, the importance of studying these systems is twofold. First, they have valuable lessons about how a scalable and fault tolerant network can be engineered. Second, their differences guide us to IoD's specific challenges which have not been tackled before and are in need of innovative solutions. We describe these structures through a discussion of goals and functionality that are relevant to IoD and the differences with IoD that need to be addressed in our architecture.

\subsection{Air Traffic Control Network}
\begin{figure}[!htb]
\centering
\includegraphics[width=2.5in]{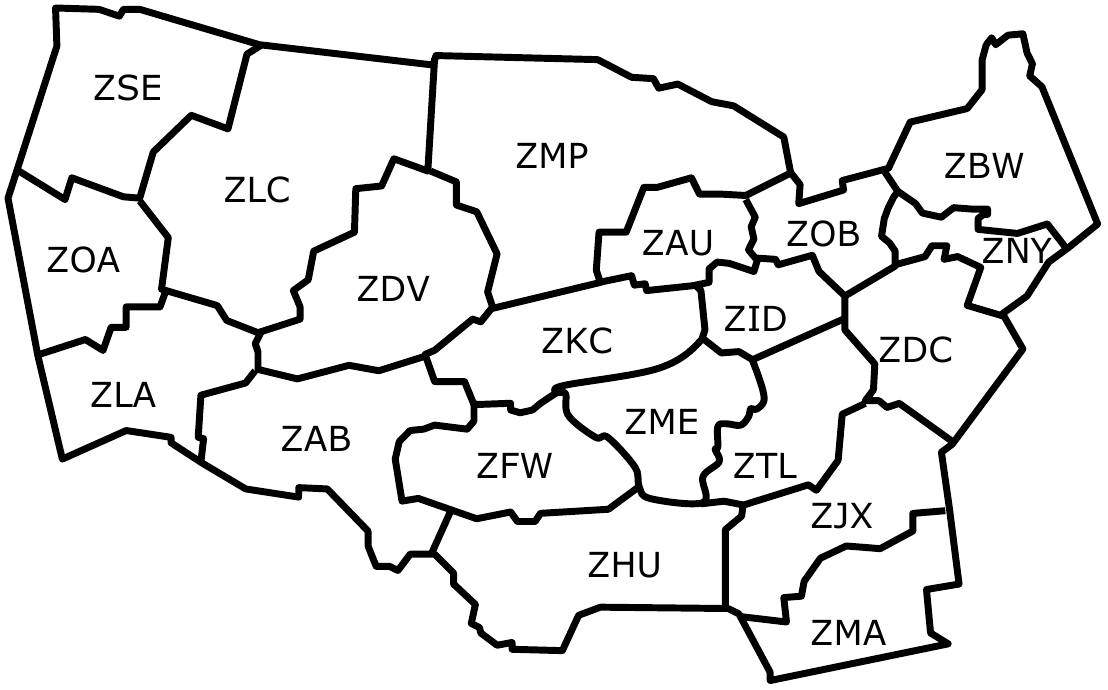}
\caption{ARTCCs in the contiguous United States (recreated from FAA\cite{FAA06}). The zones cover the airspace above and slightly beyond the contiguous United States.}
\label{FigureARTCC}
\end{figure}

ATC has strong relevance to IoD as efficiently utilizing the airspace and maintaining collision free navigation is an integral part of any IoD architecture. The functioning of ATC follows similar procedures around the globe. We briefly summarize the components of ATC in the United States. The Federal Aviation Administration (FAA) is in charge of regulations and air safety, and has partitioned the United States' airspace into 24 areas each managed by one of the 24 Air Route Traffic Control Centers (ARTCC) (Fig. \ref{FigureARTCC}). There are bilateral letters of agreement between any two adjacent ARTCCs on how aircraft must transition from one ARTCC to another. Similarly, within each ARTCC, the airspace is partitioned into between 20 to 80 sectors and each sector is exclusively managed by one controller and the aircraft transitions between sectors are done according to facility directives. The main driver in designating the boundaries of ARTCCs as well as the sectors within each ARTCC is to distribute the load in an equitable way. As it is evident in Fig. \ref{FigureARTCC}, the high volume of flights in the densely populated east coast translates into a higher number of ARTCCs than the central United States.

Traditionally, the main role of air traffic controllers was to keep a prescribed separation between all aircraft. However, within the next generation of ATC (NextGen) -- a new system with the motivation to address the lack of scalability of the current system, pilots are more autonomous and as a result in charge of their own separation and controllers intervene only when necessary. This is possible due to pilots being equipped with Automatic Dependent Surveillance-Broadcast (ADS-B) technology for navigation and localizing other aircraft in their proximity. ADS-B uses GPS for navigation and broadcasts aircraft position periodically. Use of ADS-B Out (broadcaster only with no receiver) within specific portions of airspace is mandated by 2020 \cite{GPO10}. Difficulty with aircraft localization has been a great problem in aviation, forcing most of the air traffic through certain preferred airways (analogous to the highways on the ground). However, use of ADS-B may permit what is called free flight within NextGen which allows flying in a straight line from the departure to destination airport. Unlike Internet where if some part of a network exceeds its capacity, it conveniently drops new transferring packets, this is not possible in ATC. Therefore, all ARTCC’s and sectors and airports must remain within their capacity which makes in advance reservation necessary. Flight plans are submitted to a central entity called Air Traffic Control System Command Center (ATCSCC) where according to predicted loads, a delay is assigned to each flight to ensure the network will not be oversubscribed. Pilots will receive partial or complete clearance. Once airborne, with the unfolding of how the actual flights progress, additional delays are assigned to the flights. The idea is to apply these delays as early as possible in the flight or before takeoff, rather than near the end where the maneuver space and fuel capacity are limited. These delays can be achieved by ground hold, lowering the cruising speed or by standard holding patterns. These assigned delays are communicated to the sector controllers so they know how long they must keep the aircraft in their designated sector. Interested readers are referred to \cite{Nol10} for a full treatment of air traffic control systems.

There are certain differences between IoD and ATC. As the number of drones scales up to the thousands sharing the limited airspace at any time, use of a centralized entity like ATCSCC for load prediction and assignment is not possible. Hence, one has to look at decentralized solutions. With that volume of flights, separation must be autonomously done by the drones and it is not wise to rely on human interventions for safety management, in contrast to NextGen. The limited airspace of the urban environment can only accommodate drones that have minimum performance requirement which, depending on the situation, can be stringent such as a requirement to execute holding patterns in a small area (ideally hover as in the case of vertical take off and landing (VTOL) aircraft) and ability to easily land when necessary. This opens up many possibilities within IoD for handling congestion which is not available to the ATC system. Free flight, although a step forward for ATC, is only partially implementable within IoD due to limited urban airspace, obstacles such as buildings and birds and high level of congestion anticipated. In other words, the airspace must be highly regulated to ensure smooth air traffic flow is achieved. 

\subsection{Cellular Network}
\begin{figure}[!htb]
\centering
\includegraphics[width=2.5in]{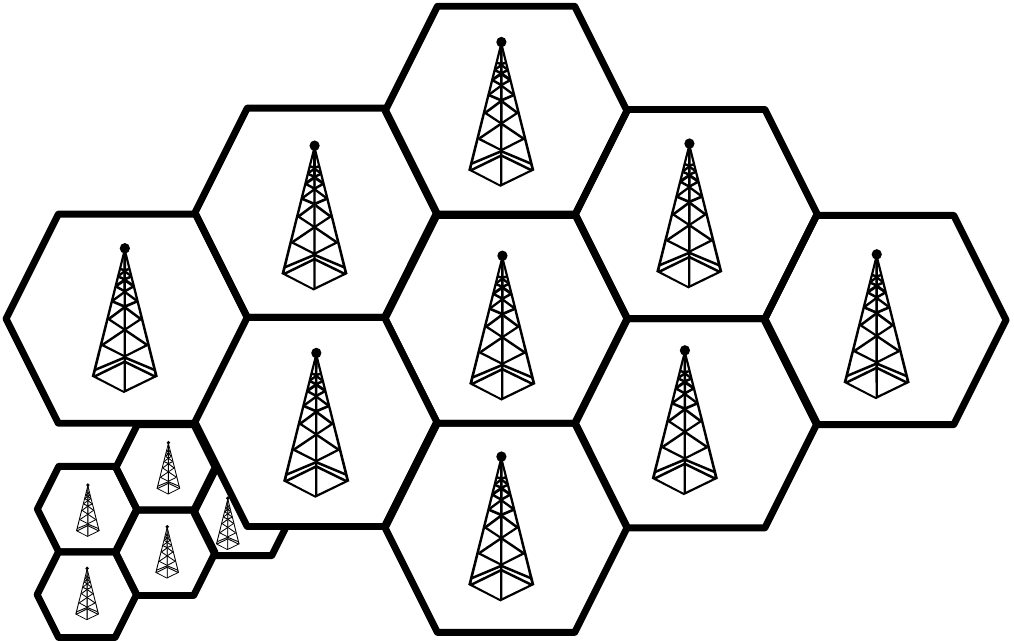}
\caption{An illustration of cellular networks and base stations.}
\label{FigureCellular}
\end{figure}

In the cellular network, the coverage area is partitioned into most commonly hexagonal cells forming a honeycomb pattern. The communication signals in each cell are sent to and received from the mobile users by a dedicated base station. Each base station uses a certain frequency which is different from the near base stations' frequencies to minimize the interference. The range of signal for each base station determines the size of each cell. Each base station can only carry a certain amount of calls over its frequency channel. As such, the main driver in determining the size of each cell is the expected number of mobile users in the region (Fig. \ref{FigureCellular}). Hence the densely populated downtown areas can have many smaller cells whereas in the rural areas, fewer cells with higher range are used. Each of the base stations are connected to a central entity called Mobile Telecommunications Switching Office (MTSO). The MTSO is in charge of periodic localization of the mobile units and assigning a base station to them. Furthermore, it assigns  channels to each call and performs the task of handoff or handover which is basically the transfer of responsibility for a moving mobile unit from one base station to the other base station as it enters a new partition. We will later use the same word in the context of IoD. See \cite{Rap96}, \cite{Gol05}, and \cite{Sta05} for a comprehensive treatment of the cellular network.

Compared to the best effort philosophy of the Internet, in telecommunication, the philosophy is that a call must not be admitted if there are not enough resources to sustain it until its completion. Hence, the handoff process poses a unique challenge as it is not known whether admitting a call in a cell will result in later termination as the mobile unit enters a new cell due to a lack of available channels in the new cell. Since the base stations usually belong to one corporation, the MTSO centrally makes decisions whether to allow access to a user in an effort to minimize the probability of a dropped call. As we will see, a similar problem exists for drones in IoD. It is much less expensive to hold a drone on the ground than to allow it to takeoff and later \textit{ground} it (order it to land) or \textit{hold} it (order to hover or execute holding patterns) due to a lack of resources. Hence, IoD has a design philosophy that is similar to that of cellular telecommunications networks. 

There are still various differences between IoD and cellular networks in an abstract level. A subtle difference is that in the case of cellular network, the MTSO does not know which cell will be the next cell a mobile unit will enter after admitting the call in the first place. But in IoD the source and destination is known to a greater extent for a trip by the drones which will allow a more optimized utilization of the network resources. Another difference is the central role MTSO plays which is in part possible because each company holds exclusive rights to certain bandwidths in the frequency spectrum. There are at least two reasons why a central design does not seem a good choice for IoD. Firstly, the tasks of IoD are computationally intensive. Hence we have to offload it to many autonomous systems which coordinate with each other. This way we reduce the complexity of the problem while settling for a less optimized solution. Secondly, as mentioned in the cellular network, a portion of the frequency spectrum is allocated exclusively to a corporation which means it has total control over its use. However, in our design for the IoD, each portion of the airspace must be shared by all the companies serving the same airspace and hence the amount of resources available to each company is less predictable. This means flight planning is a more involved task in a trade-off for a more efficient service provider market. We believe the exclusive right to the portions of the spectrum has made the entry of new competitors to the cellular market quite difficult, effectively resulting in a market with only a limited number of providers.

\subsection{Internet}
\begin{figure}[!t]
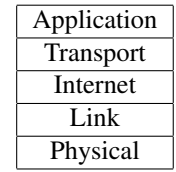

\centering
\begin{tabular}{|c|}
\hline
Application \\\hline
Transport \\\hline
Internet \\\hline
Link \\\hline
Physical \\\hline
\end{tabular}
\caption{Layers in the architecture of the Internet}
\label{FigureInternetLayers}
\end{figure}

In the Internet, the goal is to connect networks of computers together, so all the computers on the world-wide network can communicate. The Internet has a layered architecture consisting of five layers as shown in Fig. \ref{FigureInternetLayers}. Layering makes it easier to solve the problem that the Internet addresses by separating concerns. Each of these layers is to be thought as a service and upper layers use the services of lower layers. For example, the link layer is concerned solely with the transfer of data on a single communication link or between two adjacent nodes and the physical layer is concerned with the physical means for transferring signals through various mediums, such as air (in case of WiFi) or Ethernet cables. The Internet layer, relying on the connectivity provided by the link layer is concerned mainly with routing or forwarding data packets between any two nodes potentially on two different local networks through the use of standard global addressing as a best effort service rather than a reliable one. This is achieved by routers which locally make a decision about forwarding the data packets they receive to one of the immediately connected networks. Utilizing the universal unreliable connectivity provided by Internet layer, the transport layer is concerned with tasks such as the reliability of transmission and congestion control. Finally the application layer, uses this global and (if needs be) reliable connectivity for various applications like Web, Email, VoIP, Remote Login, etc. Such a decentralized and deliberately simple architecture has made the Internet a unique engineering feat in that it scaled by many orders of magnitude. Readers can refer to \cite{PD11} for a comprehensive treatment of the subject of the Internet and to \cite{BM02} and \cite{Cla88} for discussions of the philosophical guidelines in its design.

There are similarities and differences between the Internet and IoD. Routing is a task performed by both networks. However, the time scale on which the Internet operates is much smaller. In the case of IoD, the longer computation time can allow for the calculation of more optimal routes. Thus, a possibility is to adopt the routing protocols and adjust them accordingly. Another difference is that in the Internet, packets that overload the system can be conveniently dropped since it is buffered and resending it is cheap. In the IoD case, it is not possible to drop drones since they are physical objects and the only option is to remove them from the airspace by ordering them to land and providing resources to them to execute a landing order which is an expensive task. Thus some kind of reservation has to be enforced to ensure the system operates within its capacity to remain economical and viable. Whereas ATC is not a system that scales well, the Internet is designed and shown to scale well and with the expected proliferations of drones, IoD has to be an architecture that can scale. Using the Internet's design guidelines that has afforded it such  scalability, such as a decentralized design or providing generic services with the least amount of assumption about the users of the services is monumental in IoD.

\section{Our contribution}
Our core contribution is formulating a complex and multifaceted problem and showing how in an abstract level, it is related to the vast amount of existing literature on the three existing networks, namely air traffic control, cellular network, and the Internet. We have crafted a blueprint for the implementation of an IoD system based on our IoD architecture. By comparing the challenges that IoD and each of the three named networks address in an abstract way, we have established relationships between existing solutions to the specific problems of IoD, hence creating well formulated open problems for the research community in a diverse range of fields. For instance, on an abstract level all four networks have to route physical objects or data. We have uncovered this connection and others such as congestion control, admission control, and addressing schemes. We have explained the existing strategies and made it clear what prevents a straight forward adoption of them for IoD on some of these matters.

As mentioned before, although there have been numerous announcements in the media on drone applications such as package delivery prototypes, there has been no publication on the architecture for these systems. The FAA's move to address integration of drones in the national airspace\cite{FAA13}, in response to a mandate by the US House of Representatives\cite{FAA12} reiterates that IoD is a timely architecture that addresses important questions in this arena. Although there is significant excitement in the industry, to this date, this topic has not received much attention in the academic community. IoD serves as a first step for bringing these important issues to the forefront of academic endeavours and provides the academic community with well-defined problems to tackle. Our hope is that an implementation of IoD in the next three to five years will make on-demand package delivery as well as other drone applications possible.

\section{Architecture}
In this section, we explain our architecture in more detail. The purpose of our architecture is to provide extensible generic services to a diverse range of applications, namely navigation service between any two nodes in an efficient and coordinated manner as well as other common or future services such as location aware communication. A need for navigation is the common denominator for drone applications. Serving this need will enable these applications to build on top of the services provided by the architecture. Furthermore, as drones are mobile yet tasks are local, only the local drones have to be notified of the tasks, in case a pool of worker drones rather than individual drones are responsible for performing the task. Hence, providing a mechanism for location aware communication is another common need of the applications as well as other services for which the need will become apparent in the future. Two important concepts to distinguish in our paper are that of an \textit{IoD architecture} and an \textit{IoD system}: an architecture gives abstract design and feature requirements that need to be implemented by any system that is based on that architecture whereas a system gives concrete protocols (interfaces and algorithms) that implement the features required by the architecture. Hence, it is possible to have many IoD systems all based on the same core architecture each with their own advantages and disadvantages. Obviously in any engineering project, not all architectures are viable. Accordingly, at least one working IoD system must implement an IoD architecture to prove it is viable.

\subsection{Structure}
To describe our architecture, first we need to introduce a set of concepts and explain how they are related to each other in our architecture. Words with special meanings for our architecture are italicized and they  will form a vocabulary for discussing it.

\begin{figure}[!t]
\centering
\includegraphics[width=2.5in]{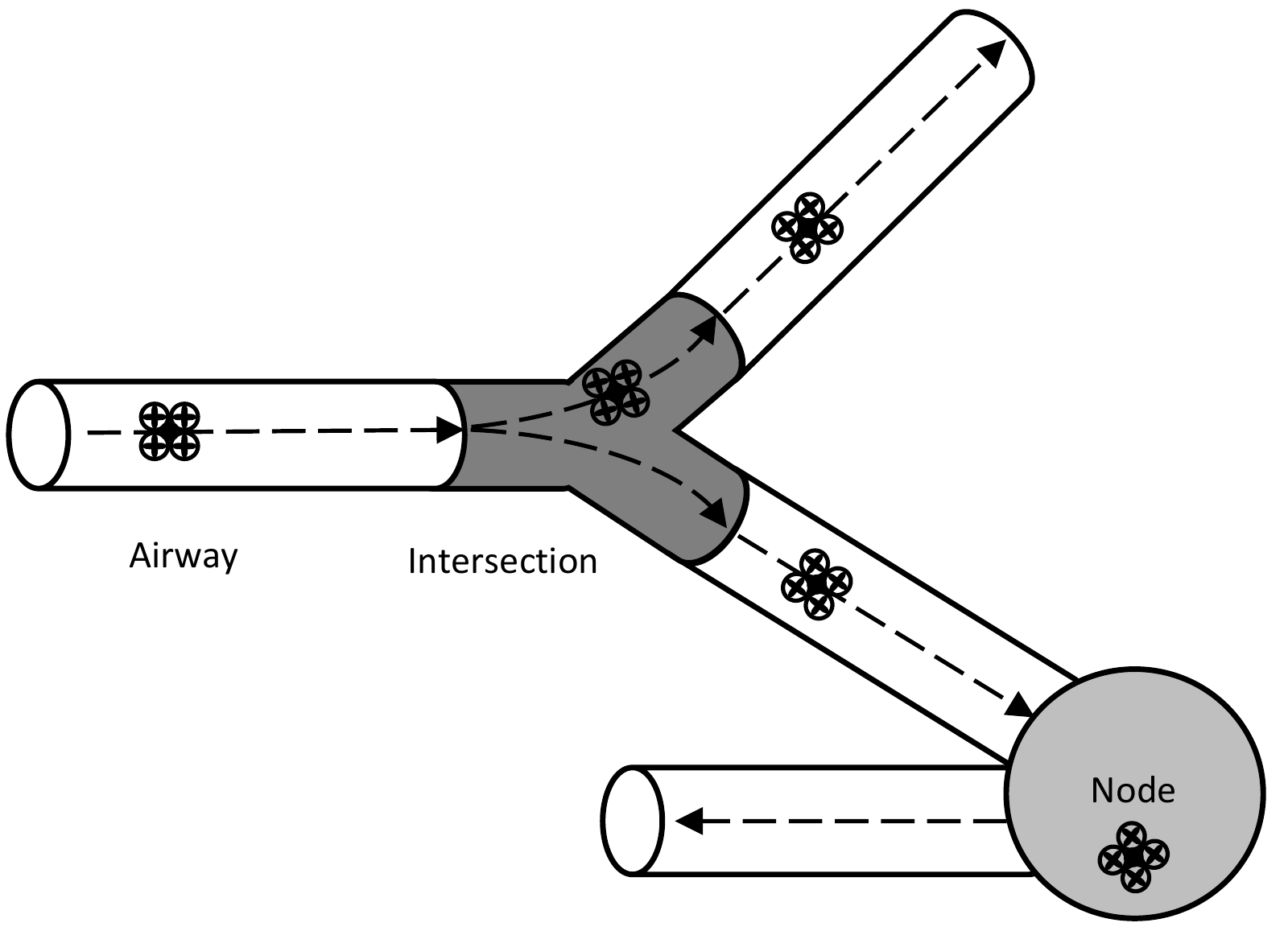}
\caption{An illustration of airways, intersections, and nodes}
\label{FigureShapes}
\end{figure}

\begin{figure}[!t]
\centering
\includegraphics[width=2.5in]{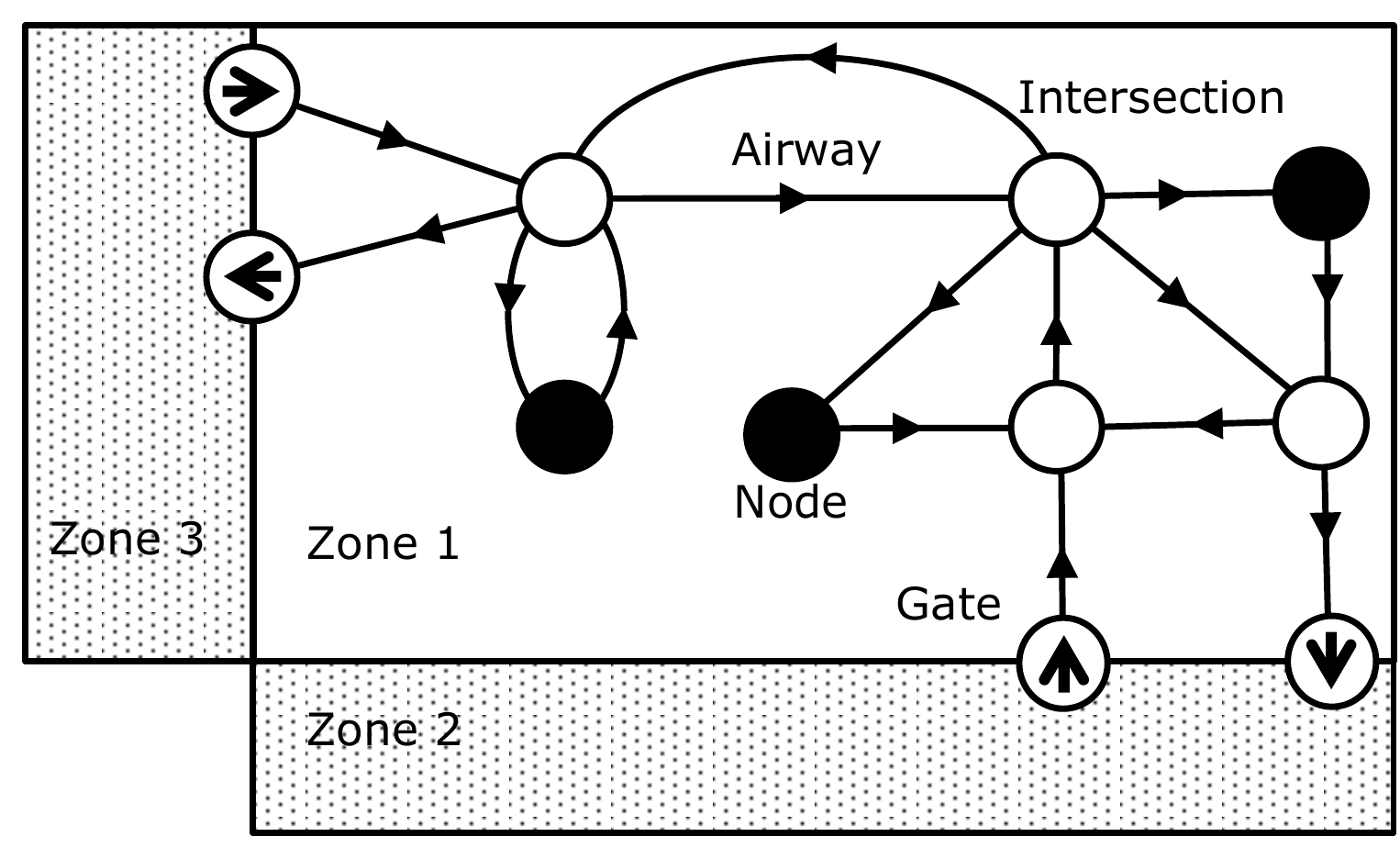}
\caption{The zone graph for zone 1 is shown. Intersections, nodes, and gates are shown with circles and marked accordingly. They constitute the vertices. Airways are shown with arrows and they are the edges of the graph. Most likely, there are many gates between any two zones, but for simplicity we show only two.}
\label{FigureZoneGraph}
\end{figure}

\begin{figure}[!t]
\centering
\includegraphics[width=2.5in]{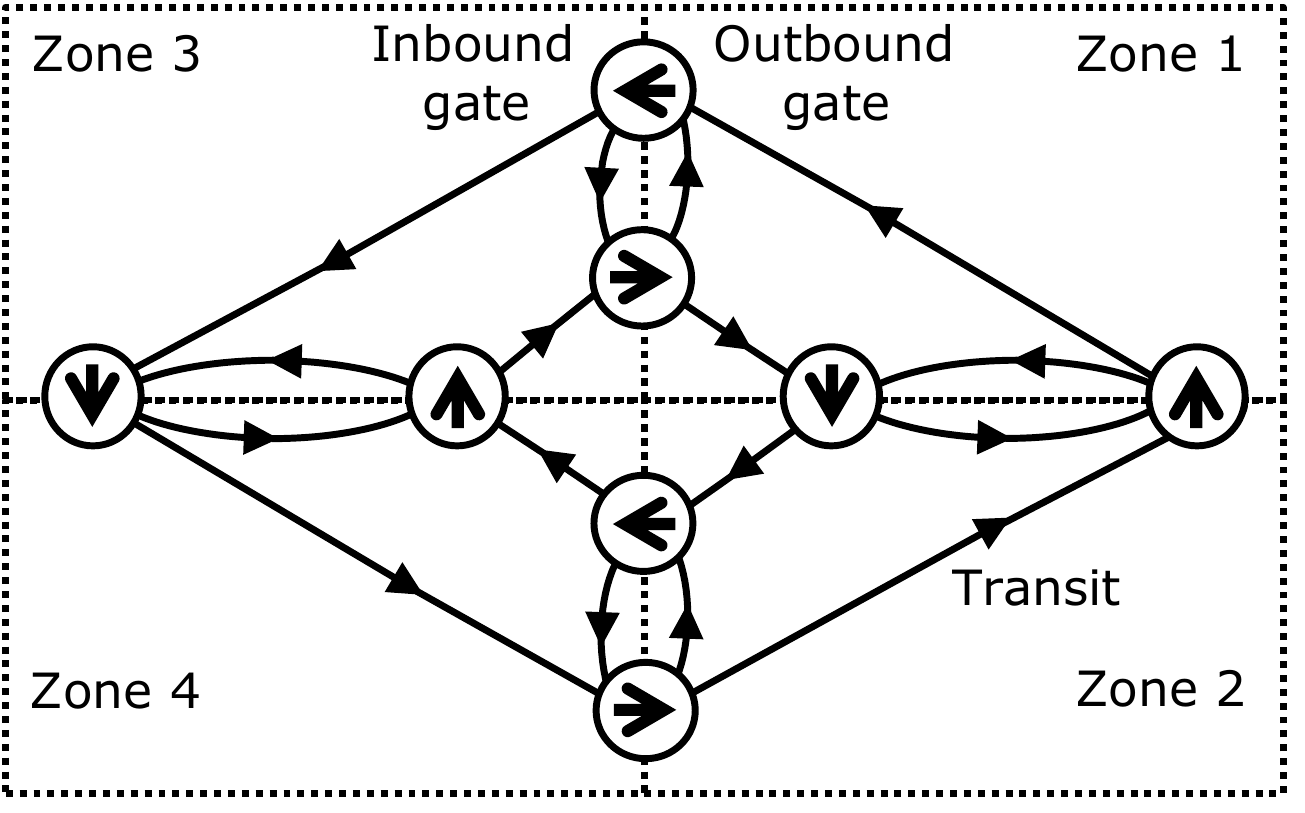}
\caption{The interzone graph for the zones 1-4 is shown. Gates are the vertices of the graph. Transits as edges are representatives of the possibility of a trip from the inbound gate to the outbound gate for the drones. Transit cost can be any cost function associated with the trip between two gates, such as average trip time. Between any two zones, there can be many gates, but for the sake of simplicity we show only two.}
\label{FigureInterZoneGraph}
\end{figure}

\textit{Airspace} is the resource that is utilized by the \textit{drones}. In our architecture, the airspace is structured similar to the roads network in the cities. Drones are only allowed inside the following three: \textit{airways} playing a similar role to the roads, \textit{intersections} formed by at least two airways, and \textit{nodes} which are the points of interest reachable through an alternating sequence of airways and intersections. Each of these three has concrete geometric shape and is guaranteed to be collision free from static structures. Movement of drones inside the airways and intersections is regulated (for example drones must move only in the designated direction(s) of an airway or intersection) whereas inside the nodes, drones are in the \textit{free flight} mode (Fig. \ref{FigureShapes}). The airspace is partitioned into \textit{zones} and hence each zone contains its airways, intersections, and nodes. Adjacent zones are reachable from each other through \textit{inbound} and \textit{outbound} \textit{gates} which are the intersections at the border but they are special in that they belong to both zones. No airway is allowed to cross the border between two zones, unless it is segmented into two airways with a gate at the border joining the airways. The graph that is formed by treating both nodes and intersections (which include gates) as the vertices and airways as the directed edges is called the \textit{zone graph} (Fig. \ref{FigureZoneGraph}). A path in the zone graph is called a \textit{pathway}. We use the word \textit{element} to refer to airways, intersections, and nodes. To be reachable, every element has a global address similar to how hosts have a global address on the Internet. If we take the gates as the vertices and connect co-zone gates with directed edges called \textit{transits}, we call the resulting graph the \textit{interzone graph}. Inside each zone, the cost of traveling between any pair of gates is called the \textit{transit cost}  where the cost can be time, distance, etc. (Fig \ref{FigureInterZoneGraph}). A path in this graph is called a \textit{route}. For the zone graph, we use the word \textit{progress} within an airway or intersection to state how far the drone has progressed the element according to some progress metric (e.g. distance from the beginning of an airway). In the zone or interzone level, the vertices and edges contain meta data e.g. in the form of \textit{components} and \textit{attributes} as in an XML tag which provide data about the particular vertex or edge. Among the meta data is the minimum \textit{performance} required from any drone that wishes to travel along the particular element, such as drone range limitations, landing restrictions, and other physical constraints. Meta data may also contain more detailed information about a particular element; for example, the meta data at a node representing a park can have a map of the park which a drone could use upon entry to the node. A portion of airspace is either \textit{public} or \textit{private}. All elements in public and private airspace are considered public and private respectively. For private elements, the \textit{access} rules for drones is specified as meta data, such as which drones are allowed access to them. At the lowest level of abstraction, we deal with \textit{points} in the airspace. The points are uniquely identified using the coordinate system of \textit{(latitude, longitude, altitude)}. For instance, an airway's geometry is understood using points. A path through points is called a \textit{trajectory}. Beware that we do not use the term trajectory in the same way it is used in robotics research where it means a time dependent path.

\subsection{Components}
\begin{figure}[!t]
\centering
\includegraphics[width=2.5in]{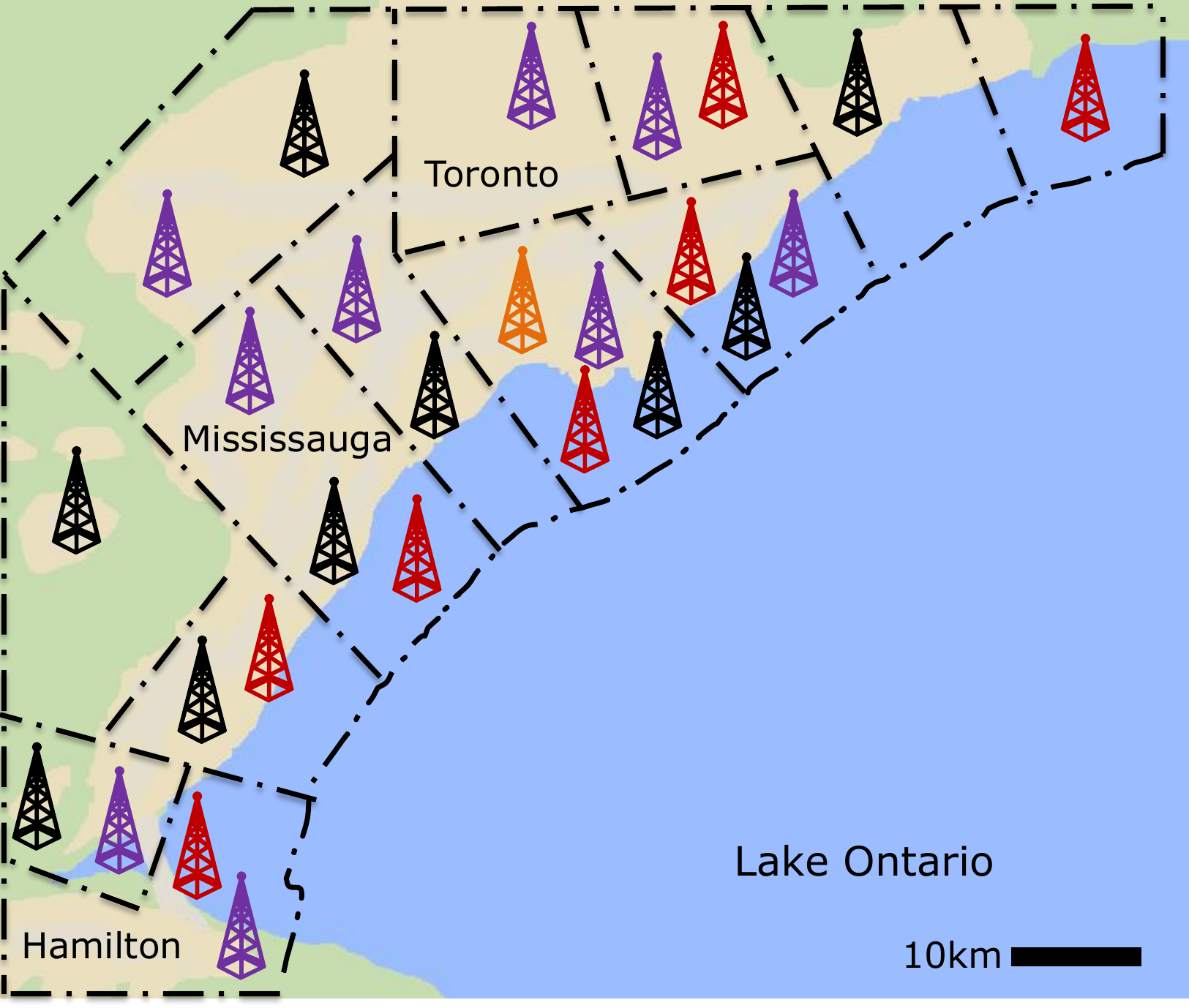}
\caption{A schematic of the Greater Toronto Area and the zones served by the ZSPs deployed by four different IoDSPs each colored differently. Handoff occurs at the boundaries of the zones.}
\label{FigureZones}
\end{figure}

Our architecture comprises of two groups of components: \textit{Zone Service Providers (ZSP)} and \textit{drones}. All ZSPs and drones are connected to the cloud, so communication between any two components is possible.
\begin{itemize}

\item[1)] In each zone, any of the ZSPs provides navigation information between any two elements 
in their designated zone to the requesting drones.  The license to operate a specific zone is granted by higher authorities. They establish and enforce the governing laws regarding the airways, intersections and public nodes such as maximum allowed drone capacity or density in them. Our architecture is not concerned with how ZSPs are realized, but it is worth mentioning that implementing a ZSP merely as software seems conceivable. We call an organization that offers ZSPs an Internet of Drone Service Provider (IoDSP). Adjacent ZSPs co-manage the gates and coordinate with each other on handoff; that is when a drone crosses the border and the responsibility has to be transferred to a new ZSP. Furthermore, ZSP can order a drone to land or hold its position by hovering or executing holding patterns and we call these actions \textit{grounding} and \textit{holding} respectively. Fig. \ref{FigureZones} presents a schematic of the Greater Toronto Area (GTA) in Canada together with the ZSPs deployed in the zones by four IoDSPs.

\item[2)] Drones are the autonomous aerial vehicles which are capable of collision free navigation along a planned route between two nodes and have various performance characteristics, such as their range, whether they are capable of VTOL and hovering, etc. They broadcast information about their position and their future path which will be used by all ZSPs, not only the particular one serving the drone. Regardless of how ZSPs and drones are implemented, they shall interact with each other through standard protocols. For instance, this allows that two competing firms have two different implementations for their ZSPs and still different drones with different implementations are able to communicate with the ZSPs through the standard protocols. Drones are required to assume fully autonomous operation beyond line of sight operation, be equipped with sense and avoid technology and be capable of emergency landing. Furthermore, specialized airworthiness certification must be considered to establish reliability levels for drones that are comparable to those of commercial aircraft when operating over inhabited areas. 
\end{itemize}


\subsection{Layers}
Similar to the Internet, we propose a layered architecture for IoD. Layering provides many benefits such as the separation of concerns, scalability, maintainability of the code base, and flexibility of modifying a layer with minimal changes needed to the other layers. The fundamental goal that the architecture is concerned with is to enable drones to perform various applications by providing common generic services for all applications. Consequently, the architecture has two goals. Firstly, it is to provide guidance to a drone from a source node to a target node and coordinate all drones' access to the airspace as a service to the drone. Secondly, it is to make available an extensible platform for other common current or future services that are needed by  applications such as delivery of messages that are intended for a pool of worker drones for an application in a specific zone (an example message is a list of local task requests). 

The navigation can be reduced to three sub-tasks. Firstly, the drone will have to traverse a path on the interzone graph from the source zone to the destination zone. Secondly, to traverse within each zone, the drone must traverse a path on the airways and intersections of the zone graph. Lastly, a trajectory of points must be chosen which the drone has to follow to stay inside the boundaries of the airways, intersections, and nodes. We tackle each of these tasks in a separate layer. The reason this seems to be a good way of tackling navigation is that having a single giant system with its map and airspace access mechanism is computationally complex and unsustainable, if not impossible. By dividing the problem into smaller sub-problems, each of them becomes more tractable. Therefore, we trade a more optimal solution for a more tractable solution. 

As mentioned, there is more needed than just navigation. For example, for a package delivery task requested by a grocery store inside some zone, only the drones that are near the store (say in the same zone) should be notified, not all the drones in the realm of IoD. Hence, ZSPs must meet these zone-specific demands through a service layer that is used by all applications. The service layer is extensible to meet the needs for future services as they will become apparent by the common needs of applications.

Our architecture consists of five layers as shown in Fig. \ref{FigureIoDLayers}. Drones have functions that fall in all the layers while ZSPs only have functions that fall under the airspace layer up to the service layer. In a strictly layered architecture, each layer provides services that are used by the layer directly above it. As is the case with the Internet (see \cite[pp. xvi,xx,xxi,33-36,87,147]{PD11}), our architecture is a relaxed layered architecture where upper layers can access lower layers and not just the layers directly below them. In effect, layering provides an effective way for logical organization of the architecture and its easy communication to other engineers and should not be treated as a never to be broken rule. The lower layers are not aware of the specifications of the higher layers. The interactions between the layers shall be through standard interfaces. The protocols then are defined between the same layers of two components.

We describe each layer in terms of the \textit{features} it is required to implement to comply with our architecture. This means that any IoD system must implement those features and define specific protocols and interfaces that make access to those features possible. We use capital letters as our convention for the name of the features.

\begin{figure}[!t]
\centering
\begin{tabular}{|c|}
\hline
Application \\\hline
Service \\\hline
End to End (E2E) \\\hline
Node to Node (N2N) \\\hline
Airspace \\\hline
\end{tabular}
\caption{Layers in the architecture of IoD}
\label{FigureIoDLayers}
\end{figure}

\subsubsection{Airspace Layer}
The airspace layer is required to implement the following features along with the needed protocols and interfaces for using these features.
\\\\MAP: ZSP is required to hold geometric representation of the elements in the zone graph; i.e. the airways, intersections, and nodes. 
\\\\AIRSPACE BROADCAST AND TRACK: Drones have to broadcast periodically their three dimensional coordinates and their future trajectories. It is conceivable these data are needed for path planning in this layer and indirectly in other layers for calculating the progress. 
\\\\PLAN TRAJECTORY: ZSP has to provide trajectories to be followed by the drone, so it stays inside the boundaries of airways, intersections, and nodes of the planned pathway. 
\\\\AIRSPACE PRECISE CONTROL: We envision a possible need for ZSP to request specific maneuvers from a drone such as holding, moving to a new point, or landing at a point. This seems to be a reasonable feature to expect from a universal architecture. 
\\\\COLLISION AVOIDANCE: In case of dynamic objects such as other drones or birds obstructing the airways or intersections, the drone must avoid colliding with them by overruling the trajectory. The drone must communicate with other drones in proximity through standard protocols for coordinated maneuvers for avoiding collision.
\\\\WEATHER CONDITION: ZSP must provide the drones with the weather conditions such as wind speed and temperature, so drones can successfully take these data into account at the time of executing a trajectory.
\\\subsubsection{Node to Node layer}
The features required for the node to node layer is as follows.
\\\\ZONE GRAPH: ZSP keeps an up to date zone graph that is augmented with the information broadcast from all the drones such as the current airway, intersection, or node of the drones and their future paths as well as their progress within an airway or intersection. ZSP knows how many drones are inside an element and roughly how they are spaced out in an airway or intersection. In the zone graph, the meta data for elements are stored too, such as the minimum performance requirement which is also a function of the weather report and changes in time. Furthermore, ZSP must provide protocols for obtaining the information in the zone graph (e.g. for viewing). Also, it must provide protocols for updating the map, such as identifying certain airways, intersection, or even the complete zone as no fly areas. Also, ZSP must provide protocols for integrating weather reports. 
\\\\N2N BROADCAST AND TRACK: Drones are required to broadcast their current element, their progress within it in case of airways or intersections, and their future path, and their estimated fuel time left periodically in a way that is accessible to all ZSPs.
\\\\PLAN PATHWAY AND CONTINGENCY: A path on the zone graph must be provided by ZSP to a drone that requests a path between any two elements as source and (intermediate) destination in the same zone. The path consists of a sequence of airways and intersections that have to be navigated for the drone to travel from the source to destination. The path does not have to be complete and a partial path for getting closer to the destination is also acceptable. A contingency path must also be provided for example to landing sites which will be used in case the drone cannot continue on its path, such as unexpected fuel shortage or when grounding by ZSP is necessary. ZSP has to take into account the performance characteristics of the drones among other things when allocating a pathway to a drone by verifying the drone meets the minimum performance rating for the paths. Also, various meta data for each element in the path can be disclosed to the drones such as the weather forecasts.
\\\\REFUEL: A path to a fuel station node (fuel station can be third party depending on the preference of the drone) should be provided by ZSP to a drone that needs to refuel. ZSP must direct the drones to fuel station nodes that are compatible with them. For example, drones can run on electricity, gas or even hydrogen (in case of fuel cell). When a drone asks for refuelling, ZSP will give a pathway to the proper fuel station accordingly.
\\\\N2N PRECISE CONTROL: It must be possible for ZSP to command the drone to hold or to move to an element, or to land at a node.
\\\\EMERGENCY: When a drone faces a software or hardware failure, if it is capable enough, it has to broadcast an SOS message to ZSP which must make arrangements such as broadcasting relevant information to all the drones, so other drones change their pathway or hold in their current element. Furthermore, ZSP must detect when a drone abruptly stops broadcasting message and issue the emergency procedures.
\\\\CONGESTION NOTIFICATION: Upon request, ZSP must provide congestion report between any two elements
inside the zone. 
\\\subsubsection{End to End layer}
The end to end layer must implement the following features.
\\\\INTERZONE GRAPH: ZSP must store the partial interzone graph at the very least. That is, it must have information at least about the the gates and transit costs in its zone (the other end of spectrum is to have complete knowledge about the interzone graph). This gives a partial or local knowledge of the interzone graph which must be learned through different means such as interaction with other zones, or input from administrator. The graph is augmented with the data broadcast from drones, so it is known which drones are inside the zone and of them which are inside the gates and which are transiting between gates. A protocol must be implemented for obtaining the data stored in the interzone graph (e.g. for viewing the graph).
\\\\ROUTING: Any two adjacent zones are likely to have several gates connecting them. The ZSPs have to provide drones with one next intermediate gate. The transit cost can be used to provide a shorter route. 
\\\\HANDOFF: Drones must be able to switch to the new ZSP when entering a new adjacent zone. ZSP must be able to handle the incoming and outgoing drones.
\\\\EXPLICIT CONGESTION NOTIFICATION: ZSP has to give explicit congestion notification on any of its gates and transits to at least the ZSPs in the adjacent zones. The algorithm to determine a gate or a transit is congested is up to the implementation by the particular ZSP.
\\\subsubsection{Service layer}
The service layer is an extensible layer that currently has the following mandatory feature and can be extended to add more services in the future as needs arise.
\\\\ZONE BROADCAST : The main role of the service layer is to provide a common platform where zone-related messages can be broadcast to the drones. For instance, a task request that needs to be performed by a drone in a particular zone can be broadcast to all the drones in that zone through service layer in ZSP. A particular task can be grocery pick up in a zone. Through encapsulation, the service layer does not understand the content of the message. However, applications by relying on the service layer for receiving the message will make sense of it.
\\\subsubsection{Application layer}
There is no feature requirement for the application layer. These are the applications that will be written in the future to use the architecture. The point of having a general airspace navigation and control service along with other services as is provided by the four layers of airspace, N2N, E2E and service is that many application we can conceive of will use these services as a foundation. So by providing it once, we enable the whole range of applications simultaneously, rather than providing a dedicated service to each application. 

\subsection{Cross-cutting features}
Any feature listed here cannot be addressed by one single layer and needs to be implemented in several layers.
\\\\SECURITY: There are a variety of threats that must be safeguarded against, among them are authentication of drones and ZSPs and other components outside the IoD system, jamming of the broadcast messages, clogging the airspace, and hacking of the drones or ZSPs. 

\section{Operation model}
Our architecture can lend itself to various operation models. We discuss one seemingly reasonable model here and in remainder of the paper we assume that we have adopted this model.

\subsection{Model}
Public is the owner of the most of airspace. There are two groups of drone owners. The first group are companies operating fleets of drones and offering various services such as logistics to users. The second group are individuals with their private drones. Since airspace is a public space, all drones are required to be registered with the government for a license to operate. Interestingly, at the time of writing this paper (December 16, 2015), FAA published an interim final rule (for a definition, see \cite{OFR11}) that mandates owners of drones with a weight between 250 grams to 25 kg to register it with the U.S. Department of Transportation\cite{FAA15-5}.

The map of where zones are located and the public airways, intersections, and nodes inside each zone is created by the municipalities in consultation with FAA as it is the ultimate aviation authority. Drone operations must be confined to inside of these elements and this must be enforced by the police. Furthermore, unauthorized entry of any drone to the private airways, intersections, and nodes is considered trespassing. Areas that do not fall into any zones are considered unregulated.

Private airspace can be defined in various ways. For example, it can be the airspace directly above a private property and below some elevation level. The municipalities set the boundaries of the private airspaces. As noted before, within private airspace, private elements are located. The owner of the private airspace, if inclined, has to design his/her own map of these elements, according to the constraints set by municipalities. The map is submitted to the municipality for the purpose of integration with the city's map along with consents for releasing the map to one, two or even all IoDSPs. Therefore, one possibility is that a private node be served exclusively by a single IoDSP, a model similar to how a host is connected to the Internet using only one ISP. At the same time any drone company can serve the node so that all the drones are potentially available to the customer resulting in faster service time. Alternatively, all IoDSPs could provide the same services to every nodes and  differentiate themselves through better implementations of protocols.

The non-exclusive license for IoDSPs to provide their services within each zone is granted by the municipalities. Airways, intersections, and nodes have to be used according to the policies set by the municipalities such as the maximum drone capacity or density. IoDSPs are obliged to provide service to all drones without discrimination (For example, an IoDSP cannot deny service to a drone in retaliation to the drone using a different IoDSP in the previous zone). These policies must be enforced by the municipalities and the police. More than one IoDSP can operate within the same zone. There is no lower or upper limit on the number of zones within which a company can operate. Any IoDSP can serve any node, as long as the private owner of the airspace has pre-authorized its access to the map through municipalities as mentioned above. And finally, during handoff (i.e. when the drone enters a new zone and the responsibility must be transferred to a new ZSP), drones can choose any ZSP in the new zone.

\subsection{Interactions with outside}
The interactions with outside are orthogonal to our architecture. To decide what entities will use the IoD system and what protocols will be used is a design choice to be made at the time of implementation. However we mention some of the entities that in all likelihood will interact with the system to give a real world picture of how an IoD system might operate.

\begin{figure}[!t]
\centering
\includegraphics[width=2.5in]{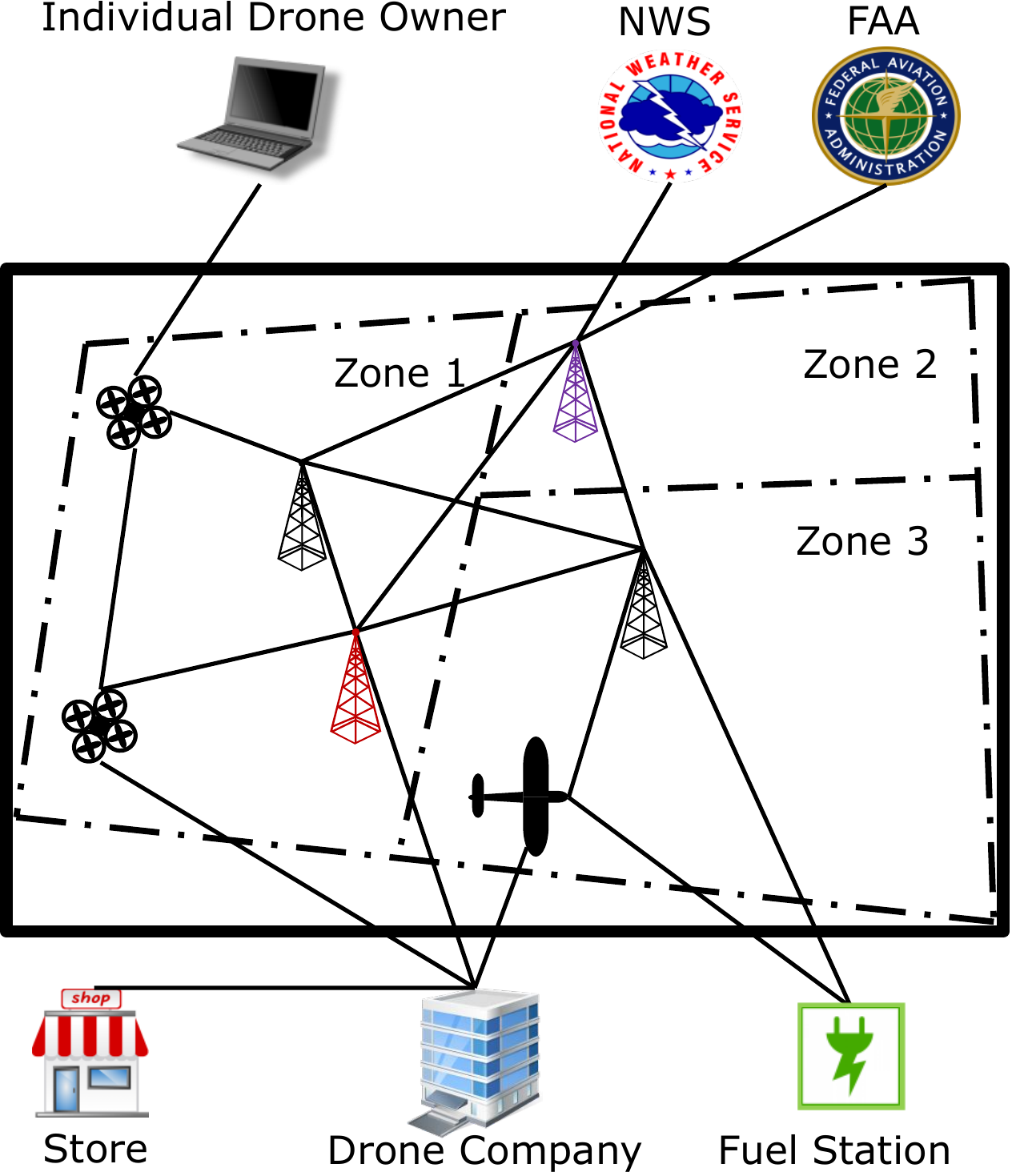}
\caption{Drones and ZSPs are components inside the boundaries of the IoD system as depicted by the box. Outside components such as fuel stations, private or corporate drone owners, governmental organizations such as weather services or FAA interact with drones or ZSPs through standard protocols. Solid lines show some of the possible interactions.}
\label{Management}
\end{figure}

An example IoD system implements protocols between the ZSPs and US National Weather Services (NWS) to disable and enable parts of the network in an automated way. The US Federal Aviation Administration (FAA) might declare a no-fly zone which is communicated through another protocol to the ZSPs. Other important entities are possibly third party fuel stations. There can be well-defined protocols for negotiating between drones and third party fuel stations (if the drone opts for using them) with the possible role of ZSPs for brokering the messages. Also, for direct messages between a fuel station and drone at the time of docking, machine to machine (M2M) protocols can be used due to the low latency that is required for the task. Since both drones and ZSPs are connected to the cloud, users, companies owning the drones, administrators, retailers like grocery stores, etc., can communicate with them through standard protocols like http (Fig. \ref{Management}). 

\subsection{Strategies for deployment}
A particularly attractive deployment strategy is the use of already deployed cellular networks. As explained earlier, in the cellular network, each provider partitions its coverage area into cells and places base stations in each of the zones. Since these base stations are already deployed, the physical space is available and they are capable of running the ZSP software. Therefore, they seem well positioned to implement ZSPs and provide wide network coverage for IoD. This strategy becomes even more interesting considering that drones have to use mobile communication which is basically what the base stations provide. This means that ZSPs will provide not only navigation, but the main communication channel for the drones. Since base stations are connected to the cloud, ZSPs can communicate with each other or with other outside entities over the cloud.

\section{Discussion and Future Work}
\subsection{Goals, principles, and benefits of our design}
In our design of the architecture, we have encouraged principles of openness, modularity, and interoperability. To achieve this, we require drones to broadcast their information using standard open protocols to communicate with ZSPs or other drones. Similarly we require interaction between all ZSPs through standard protocols. We believe it is not common or reasonable to expect competing ZSPs share traffic information and other statistics. Hence, by requiring drones to broadcast their current position and future path, we give all ZSPs in the same zone a chance to manage the traffic and have the big picture of the zone. 

One immediate benefit of openness, modularity, and interoperability is the lower overall cost for creating the navigation network. Similarly, there will be lower initial investment and lower complexity for the new IoDSPs or drone companies to enter the market. This leads to the organic growth of the network by lowering the barriers to entry and both new and existing firms will benefit from the network effect. A key consequence of these principles is that through innovation, companies will compete in their implementations while they coordinate on the standards. In our architecture, we have tried to require a minimal set of functionalities from IoD systems. Our intention is that this leaves the door open for introduction of innovative protocols and algorithms rather than the ones forced by us.

One important aspect in our design is scalability and survivability. From the experience of the Internet, survivability is a prerequisite for scalability. In an expansive IoD system, failures will be common place and the IoD system must gracefully survive them. The IoD architecture comprises of many autonomous subsystems that interact with each other only locally. This makes it possible to contain failures as opposed to have them ripple through the entire system and make it unstable. For example, the design of our architecture promotes that only a small portion of a drone's trip be reserved at any time by a local ZSP, since no ZSP has the authority to reserve a path beyond its zone. On the contrary, if the entire trip was reserved and for some reason the drone could not meet its reservation, this would affect the whole system. Furthermore, by relying on autonomous subsystems, the complexity becomes manageable as the size of computational problems that need to be solved will be substantially smaller.

An important goal in our design is to provide generic services which can serve many diverse applications. Furthermore, applications that are not even conceived today are more likely to build on top of generic services than highly specialized ones.

The concept of collision free network of airways and intersections let us circumvent the high cost of 3D mapping of the terrain and the buildings in a city. Basically, instead of guaranteeing a general statement that every possible trajectory in an area is free of obstacles (or even worse not promise that but require drones to avoid it as the only safety measure), we guarantee one example trajectory in the area that is free of obstacles (similar to the road networks). This is a substantially easier task, as mathematicians can attest to when proving a theorem versus providing an example for which the theorem holds. In addition, this enables a higher control over where drones can and cannot be (such as near airports) which is important for safety and security, and noise control (such as near hospitals or residential units). Law enforcement will be possible when drones only operate through predicted routes, as a course violation (such as trespassing in the private airspace) will be easily detected. Furthermore, this results in a more predictable traffic model which provides more organized data for planning, traffic management, and scheduling. It is worth mentioning that in ATC, a standard model for avoiding airborne collisions is the vertical separation of traffic according to direction of flights. In IoD, the model is that each airway and intersection has a specific direction of flight which is meant to achieve the same purpose.

Due to the extreme scarcity of urban airspace and safety critical nature of drone operations, we believe it is necessary to highly regulate its use through a model such as collision free network of airways and intersections as alluded to above. As such, notions like free flight as described in the context of NextGen are unlikely to be practical at least in the urban environment. At the same time we do not take the freedom away completely as free flight is possible inside the nodes by default with the extra flexibility for having collision free maps and other information if needed as the meta-data for the nodes. Nodes can have any geometries which means they have no restriction on the size. Once the drone enters a node, the ZSP no longer provides a specific trajectory as in the case of airways or intersections. ZSP will only provide the meta data about that particular node to the drone. Hence a farming drone is in the free flight mode inside a node representing a farm. If on the other hand, ZSP's help with navigation is needed for a node representing a large national park, the node must be divided into multiple nodes each connected through a network of airways and intersections, but again inside the newly formed nodes, free flight is the mode of operation. 


Each drone in an IoD system will be capable of performing one or more applications. In the Internet a user demanding a service accesses a specific host on the network and interacts with a specific application through unique IP and port numbers. However, in the IoD the dominant model is not to make requests to a specific drone directly, rather a pool of drones be ready to accept these tasks broadcast by the ZSP's service layer. This is analogous to the position of the information-centric networking line of research for the design of the future Internet (see \cite{ADI+12} and \cite{XVS14} for recent surveys). The position is that users are mainly interested in the information rather than the host to host connectivity. In our case, we advocate that users are mainly interested in a service such as package delivery or power lines inspection, not the particular drone that performs it or the particular path within which the drone travels to perform the task.

\subsection{Routing}
A fundamental question about how we implement routing is whether we reserve the entire path before the start of the trip such as e.g. in Asynchronous Transfer Mode (ATM) network technologies (see \cite{Onv94} for instance) or metaphorically we start moving while asking (from the ZSPs) for direction. The latter is the approach the Internet takes and we believe this is the superior approach for IoD systems. The main reason is that since drones take a long time to complete their trips, reserving the entire path for them is a wasteful use of the airspace as precise prediction of the future position of a drone in a complex system like IoD is not possible. As a result, in the IoD systems that we advocate, there is no guarantee a drone will complete its trip without being occasionally \textit{grounded} by ZSPs a few times along its path, due to a lack of enough airspace. However, grounding a drone is expensive in terms of energy consumption, travel delay and waste of airspace. The situation is somewhat similar to the cellular network. In both, it takes a long time for a cellphone user or a drone to enter a new cell or zone, respectively. The idea in the cellular network is that it is best to not admit a call, if it has to be dropped later. A similar policy in an IoD system is useful as it is more expensive to ground a drone than to not let it get airborne in the first place. This subject is studied extensively in the area of call admission control (CAC) (See \cite{GB06} for a survey). However, there are three major differences: 
\begin{itemize}
\item[1)] In the cellular network, the scheme needed for reserving resources is simpler. Basically one has to ensure that future cells have enough capacity to admit the mobile unit. However, in the case of reserving the zone-graph's elements for the drone, there is more than one way a drone can travel between any two nodes and hence there is more complexity in deciding whether enough resources are set aside for a particular drone or not.
\item[2)] On the other hand, in the IoD setting, if nothing unexpected happens, the path a drone will take can be partially or completely known (depending on the particular implementation of IoD system) whereas in the cellular network it is often not known to which adjacent cell the mobile unit will enter next. Hence, in the IoD, there is less uncertainty over the path.
\item[3)] The CAC decisions are made centrally, partially motivated by the fact that in the cellular network, the adjacent cells mostly belong to the same company. However, Distributed Call Admission Control (DCAC) is a possibility, as shown by the seminal papers \cite{LAN97} and \cite{NS96}, where  reservations must be made not only in the current cell of a mobile user, but also to a less extent in the neighbouring cells and the cells beyond to accommodate the mobile unit as it enters the new cells. If such a reservation is deemed possible after the base stations communicated with each other, then the call will be admitted. Referred to as the shadow clusters concept, it is similar to a quantum wave function which maps the probability of finding an electron in any region in the space where electron is analogous to the mobile unit.

\end{itemize}
It is conceivable that a similar idea for IoD inspired by DCAC can provide a routing algorithm that grounds very few drones while utilizing the airspace in an efficient way, by not reserving the entire path from the source node to the destination node for them. Otherwise, prediction errors will ripple through the whole system and make it unstable \cite{BM02}. As noted above, the routing task will be harder in IoD because of the complex structure of the resources, but the lower uncertainty over the drone's path can be useful. Developing such a routing algorithm is an important contribution to IoD.

\subsection{Congestion control}
With the possibility of thousands of drones at flight at any point in time in an urban environment, a main purpose of the IoD architecture is to coordinate access to the airspace. It is instructive to first discuss how the congestion control in the Internet works. The goal in the Internet is to ensure efficient and fair use of bandwidth. There is no central mechanism that in the short run allocates bandwidth to each of the hosts, i.e. the end nodes which are the users of the network. Rather, hosts allocate a fair and efficient amount of bandwidth to themselves in a participatory fashion. They do this by probing the network and refraining to add more loads to it if they realize the network is in a congested state. This is done by analyzing the amount of time it takes for the delivery acknowledgment (ACK) to be received by the sender (if ever in case of a dropped packet). To probe the network in a decentralized way, the network is driven toward congestion which creates delayed or lost packets which results in delayed or unsent ACKs respectively. From this, the sender realizes it must slow down in sending more packet until the network becomes less congested. This is an implicit way of inferring congestion. Today, some of the routers in the middle of the Internet are capable of sending Explicit Congestion Notification (ECN) \cite{RF99} by looking at the number of packets that they have in the queue that are not yet sent. This is a helpful feature, because running a network in a congested mode is not efficient; something that we have to do when the network does not provide feedback, just to be able to implicitly infer congestion. 

The congestion status must be known in IoD within each of ZSPs for two reasons. First, running a congested airspace translates into grounding or holding which are both expensive operations. Second, in the Internet, the ACKs happen on the orders of few hundreds of milliseconds. This fast feedback loop allows implicit congestion probing as a viable option. In IoD, probing directly with drones (i.e. by seeing if drones get stuck in the congestion or not) is orders of magnitude slower. Because of especially high cost of congestion for the IoD, we believe we have to require a feature in the IoD architecture for explicit congestion notifications to at least the neighbouring zones in the E2E level. This is not needed in the N2N level, as any ZSP has complete knowledge of the congestion on all the airways, intersections, and nodes due to the broadcasts from the drones. A major difference with ATC is that there is no central controller for the whole network (ATCSCC) that regulates the load on the whole network while each ARTCC only ensures separation, which would have a negative effect on scalability. 

In the design of a congestion control algorithm, it is an open research question how to achieve a fair and efficient allocation of the airspace while not overloading any of the elements. A mechanism that exists in the Internet literature to avoid overloading a link is a token bucket scheme (for example see \cite{SW97}) in which tokens simply represent resources and each party is given a token, only if there is a token left. However, it is not clear how such a mechanism would work for an IoD system as there are more than one ZSPs which can grant access to the same element, and being competitors, it is reasonable to assume they will not share information.

An IoD system must achieve fairness in allocation of the airspace. However, fairness is a subjective term and can lead to different designs depending on how the fairness is defined. Should we give more priority to the faster drones at the expense of slower ones, since they use the airspace for a shorter period or should we allocate the airspace to each drone in an equitable way? In IoD, similar to the Internet, related to the question of fairness is a design that takes into account the Quality of Service (QoS), i.e. the network performance according to various metrics. The interesting fact is that not all the applications have the same needs. For example, a drone that surveys the traffic has to stay aloft for extended time where short interruptions are not necessarily important whereas a drone that delivers a package needs the airspace for a short period of time and has to minimize its delivery time to meet customers' demands. 
\subsection{Communication signalling}
Since drones are wireless and ZSPs have to broadcast, there will be a high amount of communication signalling which can flood the allocated frequency channels. IoD protocols must be designed with respect to  the channel capacities as well as the number of drones and ZSP that will use the channel. If a high signalling overhead is inevitable for the functioning of IoD, then communication channels must also be treated as a resource similar to how airways and intersection are treated. Therefore, for the purpose of reserving the airspace for the drone, communication channels should be reserved as well and if any of these resources are not available a reservation should be deemed not possible.
\subsection{Addressing schemes}
Similar to the zone graph elements, drones are in need of global addressing. Whereas airways, intersections, and nodes as well as ZSPs are stationary, drones are mobile. Hierarchical addressing schemes similar to telephone numbers or IP addresses can prove useful for the zone graph elements or ZSPs. However, a particular shortcoming of the current Internet is that when IP was designed, it was assumed that it will work with stationary units. However, with the proliferation of mobile devices, that assumption is no longer valid. It seemed reasonable at the time the Internet was designed to have IP address serve two purposes; i.e. identification and localization. Identification is achieved by requiring every host to have a unique IP address. Localization is achieved by separating the IP addresses into a network portion and a host portion where each network can be part of a larger network; an idea referred to as subnetting. This design choice results in poor performance when the hosts are mobile\cite{PD11}. Therefore, any addressing scheme for drones should perhaps separate these two functions in some form as is the case with most solutions to mobility on the Internet including Mobile IP and IPv6. A particularly interesting choice would be geographical addressing\cite{SK08} where each drone is assigned an evolving address according to its current geographical position. For instance, this can provide a finer control over which drones to dispatch for a local task in a zone. 

\subsection{Drones and minimum performance}
In an IoD system, an important ability that might be mandated by authorities in high traffic areas such as lower altitude in the urban airspace is the VTOL ability which enables easier grounding or holding by hovering. This can mean that most of the urban airways, intersections, and nodes in the lower altitude may require VTOL whereas in higher urban altitude it may not be required. This is because VTOL drones are highly versatile and can perform tasks in an environment with very little airspace available to them. Most commercial aircraft are each equipped with on-board systems like Traffic Alert and Collision Avoidance System (TCAS) which are designed to decrease the chance of mid-air collision (see \cite{FAA11} for instance). It is a complex system and just to avoid collision between two aircraft, thousands of lines of code are needed. In our case of lower altitude urban airspace, it is reasonable to assume that in often congested area with thousands of drones in flight, to avoid mid-air-collision, aircraft must be able to hover and move vertically to regulate the traffic, similar to the road network and cars which can stop. Drones are ultimately responsible for avoiding collisions mid-air and a TCAS like system for drones without hovering abilities is a major challenge for more than two drones.

\subsection{Security}
Security is not a topic that can be addressed by any single layer. A major challenge in the Internet today is that security is mostly provided by the application layer and there is a lack of in-place security mechanism in the lower layers. The Internet has been exploited for its security vulnerabilities which have led some network researchers to consider the security as one of the main goals in the next architectures for the Internet \cite{Fel07,Rob09,BCPS06}. Arguably, damages from malicious users are more severe in the case of IoD compared to the Internet and security must be one of the core issues that any architecture for IoD should address. Given the experience from the Internet,  we required in our architecture that security be implemented across all the layers, as it is a cross-cutting concern.

\subsection{Validation and technical implementation}
This paper presents a conceptual architecture and the grand technical contribution is to instantiate at least one system based on it to validate and demonstrate that our architecture can work in practice. This entails designing protocol suites and interfaces between the layers and implementing the layers with the required features. Any inconsistency or inefficiency revealed at the time of implementation can be used for later iterations of the architecture. To implement an IoD system there are many non-programming questions that have to be answered, such as the questions discussed about routing and congestion control. This will be the main area of our focus in the future works. Building IoD is a great undertaking which needs the participation of the research community at large. By presenting the architecture in the current stage, useful protocols can be discussed and designed by the research community which can be validated once a simulation as well as a physical platform for IoD is ready. Furthermore, the design of the IoD architecture itself can benefit from the work of the researchers working on diverse range of networks from air traffic control to cellular to the Internet who will apply their knowledge to IoD.

\subsection{Economics of IoD}
From an economic point of view, the operation model and the protocols of the system must provide enough incentives to the stakeholders to pursue the desired actions. It is interesting to study related questions through the lens of game theory and mechanism design.

\subsection{Legislation}
Another major topic is to provide a legal framework for the IoD. One of the main barriers in utilizing the drones today is lack of legislation that properly address the technology. This is manifested in the recent  Public Law 112-95 titled ``FAA modernization and reform act of 2012''\cite{FAA12} enacted by US House of Representatives. In the Public Law 112-95, the secretary of Transportation is mandated among other things firstly to develop a comprehensive plan for expediting the integration of civil Unmanned Aircraft Systems (UAS) into the national airspace system, and secondly create a 5 year roadmap for their introduction. Thirdly, specifically for small unmanned aircraft system, a rulemaking was required that would expedite the start of their civilian operation in national airspace system.

In response, in 2013, FAA along with other governmental agencies jointly published a comprehensive plan for integration of UAS into national airspace system\cite{FAA13}. In this document, UAS national goals and objectives are described. One of the goals is to make civil visual-line-of-sight operation of small UAS a routine by 2015. Initially this will be outside of class B and C airspace and above urban areas. In accordance with this goal, in February 2015, FAA published a notice of proposed rulemaking \cite{FAA15-2} that addresses introduction of small UAS (i.e. weighing less than 25kg) into national airspace system. Various safety measures have been proposed such as visual-line-of-sight operations. Flights are restricted to day time at a maximum altitude of 152.4m above the ground. Small UAS cannot operate in class A airspace. However, operation within class B,C,D, and E airspace is possible with permission from ATC. Furthermore, operation in class G airspace does not require a permission from ATC. As mentioned before, for UAS that weighs more than 250 grams, the owner has to register it with FAA \cite{FAA15-5} for outdoor operation. Another goal set forth in the comprehensive plan is to make routine operation of UAS possible in the national airspace by 2015 for the public organizations and by 2020 for the civilians.

To comply with the public law \cite{FAA12}, FAA has published a roadmap for integration of UAS in the 
national airspace system\cite{FAA13-2}. Currently for UAS to access the airspace, Certificates of Waiver or 
Authorization (COA) are needed for public operation and certain airworthiness certificates for experimental 
civil application as mandated in \cite{FAA12}. Initially FAA plans to accommodate UAS in the near-term (next 
5 years), then it transitions into the period of integration (5-10 years) in the mid-term and in the long-
term (more than 10 years) it is expected that requirements from UAS will evolve based on the safety 
requirements from all type of aircraft and is consistent with the timeline for NextGen vision. FAA asserts 
that for UAS to be allowed access to the national airspace, they must be able to apply and be accepted for 
standard airworthiness certificate.

A challenging goal for the FAA is to integrate UAS without segregating various types of aircraft.
Two important required technologies according to the FAA's roadmap is Sense and Avoid (SAA) and Control and Communication (C2). The SAA is expected to ensure self-separation and at a later stage collision avoidance which needs to be interoperable with other collision avoidance systems as well as compatible with ATC separation services. According to the FAA, third party-communication service providers are used frequently today and it is a routine task for FAA to effectively monitor their performance. The choice of the right type of third party C2 providers is dependent on the choice of UAS architecture. At International Telecommunication Union’s World Radiocommunication Conference in 2012, an agreement was reached to dedicate a part of frequency spectrum for exclusive use by UAS. This paves the way for the operation of UAS across international borders and protects UAS from interference from other devices. \cite{FAA13-2}

According to the roadmap\cite{FAA13-2}, the FAA asserts that unless new classes of airspace are specifically created for UAS, for them to be accepted for integration in the national airspace system, they must satisfy the following requirements from FAA (with notable exception of line-of-sight small UAS). In addition to airworthiness certificates alluded to above, any UAS must register and execute an IFR flight plan (see \cite{Nol10} for a definition) and be equipped with ADS-B (Out); i.e. the ADS broadcasting component. Furthermore, they have to meet the minimum performance and equipage requirement of the area where the operation takes place. Additionally, each UAS must have a flight crew including a pilot-in-command who is only in charge of only one UAV and fully autonomous operations will not be allowed. Also, minimum required separation must be met in the controlled airspace and ATC will be in charge for separation services for the applicable airspace classes for manned and unmanned aircraft.

In our opinion, the ban on the fully autonomous operation set forth by the FAA in their roadmap takes away the major benefits of any drone architecture, including IoD. Fortunately, the FAA does not rule out the introduction of new classes specifically designated for UAS in their roadmap as mentioned above. Certainly, IoD in its current form is a theoretical framework that is only viable if these new classes are introduced. According to \cite{FAA13-2}, the FAA provides a transparent process for setting regulations which encourages comments from the public as well as other feedback mechanisms for avoiding onerous regulations. In the process of crafting new legislation, the FAA has been soliciting feedback from the UAS community with one example being creation of the Advisory and Rulemaking Committees(ARC) for UAS comprising of members from industry and academia\cite{FAA11-2}. We are optimistic that the stakeholders will influence the process in a way that new airspace classes are created for UAS rather than what we believe is fitting a fundamentally new technology into a frame that was designed for a different technology.

\section{Conclusion}
Many drone applications can benefit from a unified framework that coordinates their access to the airspace and helps them navigate to the points of interest where they have to perform a task. Any architecture poised to provide this service must be scalable and be able to provide it to thousands of drones, which will share the congested and limited urban airspace. 

In this paper, we laid out the conceptual foundation for such an architecture by developing a vocabulary of concepts for describing the architecture and identifying the relevant components of it as well as deciding on the boundaries of the architecture. Furthermore, we designed a structure for the airspace and provided strategies for utilizing that structure in the airspace. Our design makes it possible to provide generic services that can be used by many applications. To effectively tackle the problem of ``how to enable drones to perform tasks'', we divided the overall required functionality of the architecture into logical layers. The main sub-problem was the airspace navigation and coordination for various applications as addressed in the first three layers of IoD. We addressed other common services that are needed by applications such as location aware communication in an extensible service layer. In IoD architecture, we describe the features that are required to be implemented in each of these layers by IoD systems. Furthermore, we suggested an operation model that identifies the role of  private and public organizations in the governance of IoD. Additionally, we explored and discussed some of the difficulties that have to be addressed for an effective IoD system. In all of this, we used and referred to the wealth of knowledge acquired from three large scale networks, the cellular network, air traffic control, and the Internet. Finally, we discussed the differences and future works that can benefit from the solutions from the vast existing literature on these three subjects.


%



\section*{Acknowledgment}
We would like to thank Yashar Ganjali for fruitful discussions on the subject.

\ifCLASSOPTIONcaptionsoff
  \newpage
\fi



\bibliographystyle{IEEEtran}
\bibliography{IEEEabrv,internetofdrones}
%





%
\newpage

\begin{IEEEbiography}[{\includegraphics[width=1in,height=1.25in,clip,keepaspectratio]{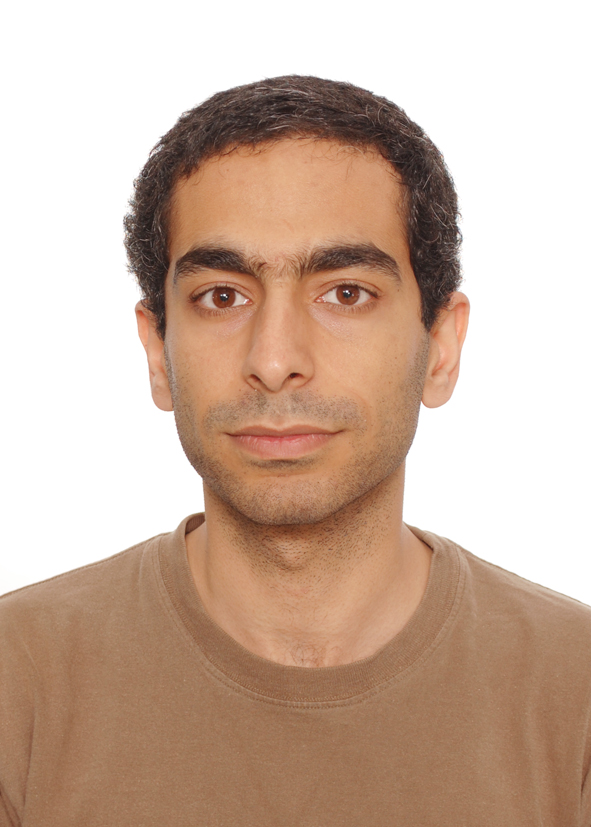}}]{Mirmojtaba Gharibi}
received the B.ASc. in electrical engineering from Sharif University of Technology. He completed his M.Math, and currently is pursuing a PhD degree both in Computer Science at University of Waterloo, Waterloo, Canada.
\end{IEEEbiography}

\begin{IEEEbiography}[{\includegraphics[width=1in,height=1.25in,clip,keepaspectratio]{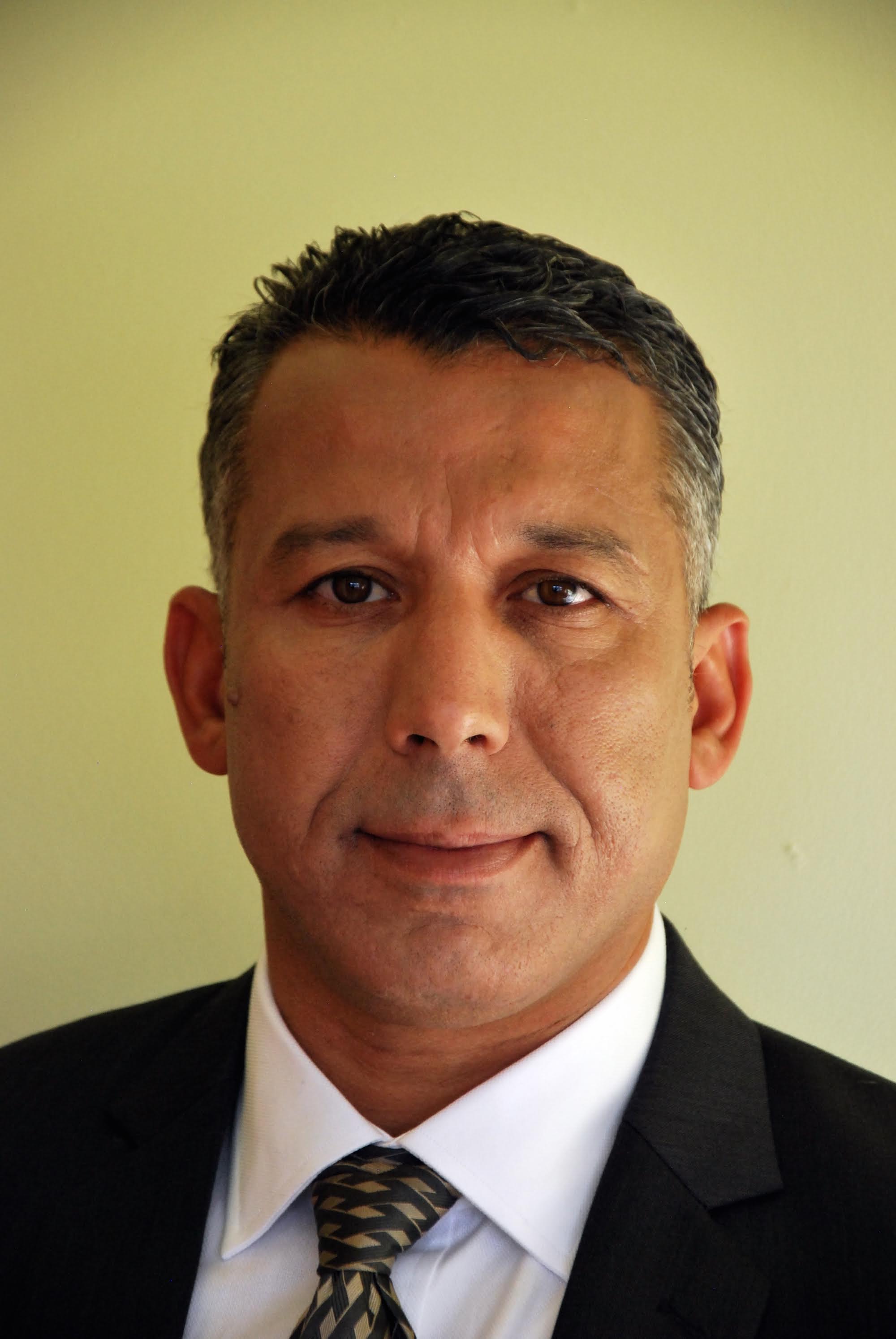}}]{Raouf Boutaba}
received the M.Sc. and Ph.D. degrees in Computer Science from the University Pierre \& Marie Curie, Paris, in 1990 and 1994, respectively. He is currently a professor of Computer Science at the University of Waterloo. His research interests include resource and service management in networks and distributed systems. He is the founding editor in chief of the IEEE Transactions on Network and Service Management (2007–2010) and on the editorial boards of several other journals. He has received several best paper awards and other recognitions such as the Premier’s Research Excellence Award, the IEEE Hal Sobol, Fred W. Ellersick, Joe LociCero, Dan Stokesbury, Salah Aidarous Awards, and the McNaughton Gold Medal. He is a fellow of the IEEE, the Engineering Institute of Canada, and the Canadian Academy of Engineering.
\end{IEEEbiography}

\begin{IEEEbiography}[{\includegraphics[width=1in,height=1.25in,clip,keepaspectratio]{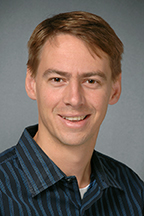}}]{Steven L. Waslander} received the B.Sc.E degree in Applied Mathematics and Mechanical Engineering from Queen’s University, Kingston, ON, Canada, in 1998, the M.S. in Aeronautics and Astronautics from Stanford University in 2002 and the Ph.D. in Aeronautics and Astronautics from Stanford University in 2007. He is currently an Associate Professor with the Department of Mechanical and Mechatronics Engineering, University of Waterloo, Waterloo, ON. He is also the Director of the Waterloo Autonomous Vehicles Laboratory. His main research interests include perception, navigation and control of autonomous aerial rotorcraft and ground rovers with a focus on simultaneous localization and mapping, optimal motion planning and multi-robot coordination.

\end{IEEEbiography}




\end{document}